\documentclass{natureprintstyle}
\usepackage{ifthen}
\usepackage{graphicx}
\usepackage{amsmath}
\usepackage{amssymb}

\newboolean{SubmittedVersion}
\setboolean{SubmittedVersion}{False}
\DeclareMathAlphabet{\mathpzc}{OT1}{pzc}{m}{it}

\def\Rb87{^{87}\text{Rb}}
\def\Na23{^{23}\text{Na}}
\def\Li6{^{6}\text{Li}}

{\catcode`\|=\active
  \gdef\Braket#1{\left<\mathcode`\|"8000\let|\BraVert {#1}\right>}}
\def\BraVert{\egroup\,\mid@vertical\,\bgroup}

\begin{document}

\title{Quantum phases in circuit QED with a superconducting qubit array}
\author{Yuanwei~Zhang$^{1,2}$, Lixian~Yu$^{2,3}$, J. -Q.~Liang$^{1} $,
Gang~Chen$^{2,*}$, Suotang~Jia$^{2,\dag}$, Franco~Nori$^{4,5,6,\ddag}$ }
\maketitle

\begin{affiliations}
$^*$Correspondence to chengang971@163.com\\
$^{\dag}$Correspondence to tjia@sxu.edu.cn\\
$^{\ddag}$Correspondence to fnori@riken.jp\\
\\
\item Institute of Theoretical Physics, Shanxi University, Taiyuan 030006, P. R.
China \\
\item  State Key Laboratory of Quantum Optics and Quantum Optics Devices, Institute of Laser spectroscopy, Shanxi University, Taiyuan 030006, P. R. China \\
\item  School of Physical Science and Technology, Soochow University, Suzhou,
Jiangsu 215006, P. R. China\\
\item  CEMS, RIKEN, Saitama 351-0198, Japan\\
\item  Physics Department, The University of Michigan, Ann Arbor, Michigan 48109-1040, USA\\
\item  Department of Physics, Korea University, Seoul 136-713, Korea
\end{affiliations}

\begin{abstract}
Circuit QED on a chip has become a powerful platform for simulating complex
many-body physics. In this report, we realize a Dicke-Ising model with an
antiferromagnetic nearest-neighbor spin-spin interaction in circuit QED with
a superconducting qubit array. We show that this system exhibits a
competition between the collective spin-photon interaction and the
antiferromagnetic nearest-neighbor spin-spin interaction, and then predict
four quantum phases, including: a paramagnetic normal phase, an
antiferromagnetic normal phase, a paramagnetic superradiant phase, and an
antiferromagnetic superradiant phase. The antiferromagnetic normal phase and
the antiferromagnetic superradiant phase are new phases in many-body quantum
optics. In the antiferromagnetic superradiant phase, both the
antiferromagnetic and superradiant orders can coexist, and thus the system
possesses $Z_{2}^{z}\otimes Z_{2}$\ symmetry. Moreover, we find an
unconventional photon signature in this phase. In future experiments, these
predicted quantum phases could be distinguished by detecting both the
mean-photon number and the magnetization. \newline
\end{abstract}


Circuit quantum electrodynamics (QED) based on superconducting qubits is a
fascinating topic in quantum optics and quantum information \cite%
{JQY03,AB04,AW04}. This artificial spin-$1/2$ particle can be controlled by
tuning the external magnetic flux and gate voltage \cite{JQY05,RJS08,JC08}.
Moreover, a strong spin-photon coupling has been achieved, which allows to
implement quantum operations for long coherence times \cite{BI11,ZLX13}.
Recently, many important quantum effects in atomic physics and quantum
optics have been observed in this artificial spin-photon interaction \cite%
{IB09,JQY11}. Particularly, experiments realized multiple superconducting
qubits interacting with a transmission-line resonator \cite{JMF09}. These
experiments allow to explore many-body phenomena via circuit QED \cite%
{JK10,YH11,AAH12,FN12,DM12,MH12,DLU12,JJ13,FN13,MF13}. For example, the
challenging Dicke quantum phase transition from a normal phase to a
superradiant phase, which was predicted more than 30 years ago \cite%
{Hepp,Wang,Hioes}, can be realized by controlling the gate voltage or
external magnetic flux \cite{GC07,NL09,PN11,HI12} and the no-go theorem
arising from the Thomas-Reich-Ruhn sum rule may be overcome \cite{PN10,OV11}%
.Moreover, the Jaynes-Cummings lattice model \cite{GAD06} can also be
simulated by an array of transmission-line resonators, each coupled to a
single artificial particle \cite{ADG07,HMJ08}. In addition, by measuring the
microwave photon signature, the many-body nonequilibrium dynamics, as well
as the known phase diagrams, could be derived \cite{YDW07,OV13,OV13-2}.

On the other experimental side, superconducting qubits can couple with each
other, forming an array with an effective nearest-neighbor spin-spin
interaction \cite{YAP03,MWJ11}. Thus, it is meaningful to explore the
many-body physics when a superconducting qubit array couples with a
transmission-line resonator because there exists a competition between the
collective spin-photon interaction and the nearest-neighbor spin-spin
interaction. Recently, sudden switchings, as well as a bistable regime
between a ferromagnetic phase and a paramagnetic phase, have been predicted
\cite{LT10}, attributed to this competition.

In this report, we investigate the quantum phases in circuit QED with a
superconducting qubit array, which is governed by a Dicke-Ising model with
an antiferromagnetic nearest-neighbor spin-spin interaction. By considering
the competition between the collective spin-photon interaction and the
antiferromagnetic nearest-neighbor spin-spin interaction, we predict four
quantum phases, including: a paramagnetic normal phase (PNP), an
antiferromagnetic normal phase (ANP), a paramagnetic superradiant phase
(PSP), and an antiferromagnetic superradiant phase (ASP). The ANP and the
ASP are new phases in many-body quantum optics. In the ASP, both the
antiferromagnetic and superradiant orders can coexist, and thus the system
possesses $Z_{2}^{z}\otimes Z_{2}$\ symmetry, i.e., both $U(1)$ and
translation symmetries are broken simultaneously. Moreover, we find an
unconventional photon signature in this phase which could increase from zero
to a finite value and then decrease when increasing an effective magnetic
field. In future experiments, these predicted quantum phases could be
identified by detecting both the mean-photon number and the magnetization.%
\newline

{\LARGE \textbf{Results}}

\textbf{System and Hamiltonian. }Figure 1 shows our proposed quantum
network. Many superconducting qubits connected in a chain couple
capacitively to their neighboring qubits and also interact identically with
a one-dimensional transmission-line resonator. The corresponding Hamiltonian
is given by \cite{JQY05,RJS08,JC08}
\begin{equation}
H_{1}=e^{2}\sum\limits_{i,j}\left( n_{i}-N_{g}\right) ^{T}\bar{C}%
_{ij}^{-1}\left( n_{j}-N_{g}\right) -\sum_{i}E_{J}\cos \varphi _{i},
\label{T1}
\end{equation}%
where $n_{i}$ is the number of Cooper pairs on the $i$th island,
\begin{equation}
N_{g}=\frac{C_{g}V_{g}}{2e}  \label{P1}
\end{equation}%
is the dimensionless gate charge with gate capacitance $C_{g}$ and gate
voltage $V_{g}$,
\begin{equation}
E_{J}=2E_{J}^{0}\cos \left( \pi \frac{\Phi _{x}}{\Phi _{0}}\right)
\label{P2}
\end{equation}%
is the tunable Josephson tunneling energy with single-junction Josephson
energy $E_{J}^{0}$, external magnetic flux $\Phi _{x}$, and magnetic flux
quantum $\Phi _{0}$, $\varphi _{i\text{ }}$is the phase of the
superconducting order parameter of the $i$th island, and $\bar{C}$ is the
capacitance matrix. The element $C_{ii}=C_{\Sigma }$ is the total
capacitance connected to the $i$th island and $C_{ii\pm 1}=-C$ is the
coupling capacitance between two adjacent superconducting qubits. In
general, the coupling capacitance $C$ is much smaller than the total
capacitance $C_{\Sigma }$. As a consequence, the next-nearest-neighbor term
of $\bar{C}$ can be neglected safely and the Hamiltonian $H_{1}$ is
rewritten as
\begin{equation}
H_{2}=\sum_{i}\left[ \frac{2e^{2}}{C_{\Sigma }}(n_{i}-N_{g})^{2}-E_{J}\cos
\varphi _{i}\right] +\frac{4e^{2}C}{C_{\Sigma }^{2}}%
\sum_{i}(n_{i}-N_{g})(n_{i+1}-N_{g}).  \label{T2}
\end{equation}%
Near the degeneracy point, only a pair of adjacent charge states ($n_{i}=0$
and $n_{i}=1$) on the island are relevant. If we define these charge states
as the effective spin basis states $\left\vert \uparrow _{i}\right\rangle $
and $\left\vert \downarrow _{i}\right\rangle $, i.e., $\left\vert
n_{i}=0\right\rangle \longleftrightarrow \left\vert \uparrow
_{i}\right\rangle $ and $\left\vert n_{i}=1\right\rangle \longleftrightarrow
\left\vert \downarrow _{i}\right\rangle $, the Hamiltonian $H_{2}$ reduces
to the form
\begin{equation}
H_{3}=J\sum_{i}\sigma _{z}^{i}\sigma _{z}^{i+1}+\varepsilon \sum_{i}\sigma
_{z}^{i}-\frac{E_{J}}{2}\sum_{i}\sigma _{x}^{i},  \label{T3}
\end{equation}%
where $\sigma _{z}^{i}=\left\vert \uparrow _{i}\right\rangle \left\langle
\uparrow _{i}\right\vert -\left\vert \downarrow _{i}\right\rangle
\left\langle \downarrow _{i}\right\vert $ and $\sigma _{x}^{i}=\left\vert
\uparrow _{i}\right\rangle \left\langle \downarrow _{i}\right\vert
+\left\vert \downarrow _{i}\right\rangle \left\langle \uparrow
_{i}\right\vert $ are the Pauli spin operators,
\begin{equation}
\varepsilon =2e^{2}\left( \frac{C}{C_{\Sigma }^{2}}+\frac{1}{C_{\Sigma }}%
\right) \left( N_{g}-\frac{1}{2}\right)  \label{P4}
\end{equation}
is an effective magnetic field, and
\begin{equation}
J=\frac{e^{2}C}{C_{\Sigma }^{2}}  \label{P5}
\end{equation}%
describes the capacitance-induced nearest-neighbor spin-spin interaction.

Now a one-dimensional transmission-line resonator is placed in parallel to
the superconducting qubits. All superconducting qubits are situated at the
antinode of the magnetic field induced by the oscillating supercurrent in
the transmission-line resonator \cite{AB04,AW04}. Due to the boundary
condition at the end of the transmission-line resonator, these
superconducting qubits are controlled only by the magnetic component, which
shifts the original magnetic flux $\Phi _{x}$ by
\begin{equation}
\tilde{\Phi}_{x}=\Phi _{x}+\frac{g_{0}\Phi _{0}}{\pi }\left( a+a^{\dag
}\right) ,  \label{Phi}
\end{equation}%
where
\begin{equation}
g_{0}=\frac{\pi S_{0}}{d\Phi _{0}}\sqrt{\frac{\hbar l\omega }{L_{0}}},
\label{go}
\end{equation}%
$l$ is the inductance per unit length, and $S_{0}$ is the enclosed area of
the superconducting qubit. In the Lamb-Dicke limit $(g_{0}\sqrt{\left\langle
a^{\dag }a\right\rangle +1}\ll 1)$, together with the condition $\Phi
_{x}=\Phi _{0}/2$, an effective Hamiltonian for Fig. 1 is obtained by
\begin{equation}
H=J\sum_{i}\sigma _{z}^{i}\sigma _{z}^{i+1}+\varepsilon \sum_{i}\sigma
_{z}^{i}+\frac{g}{\sqrt{N}}\sum_{i}\sigma _{x}^{i}\left( a+a^{\dag }\right)
+\omega a^{\dag }a,  \label{FNH}
\end{equation}%
where
\begin{equation}
g=\sqrt{N}E_{J}^{0}g_{0}  \label{P6}
\end{equation}%
is the collective spin-photon coupling strength. In this Hamiltonian, all
parameters can be controlled independently. For example, the effective
magnetic field $\varepsilon $ can be tuned via the gate voltage $V_{g}$ from
the negative to the positive. For simplicity, we address mainly the case $%
\varepsilon \geq 0$ in the following discussions.

The Hamiltonian (\ref{FNH}) is a Dicke-Ising model with an antiferromagnetic
nearest-neighbor spin-spin interaction. This Hamiltonian shows clearly that
when both $g$ and $J$ coexist, the collective spin-photon interaction has a
competition with the nearest-neighbor spin-spin interaction. As a result, it
exhibits exotic phase transitions beyond the previous predictions of the
standard Dicke (Ising) model. For example, a first-order superradiant phase
transition has been predicted \cite{CFL04,SG11}, when $J<0$. In this report,
we will find rich quantum phases including the PNP, the ANP, the PSP, and
the ASP for $J>0$.

\textbf{Quantum phases.} For the Hamiltonian (\ref{FNH}), the quantum phases
can be revealed by calculating the ground-state energy and the order
parameters via a mean-field approach \cite{SS99}. In the classical picture,
the spin in the Hamiltonian (\ref{FNH}) can be represented as a vector line
in the $xz$ plane with the unit vector $\overrightarrow{n}=\left( \cos
\varphi ,0,\sin \varphi \right) $. Thus, we can introduce a variational
ground-state wave function
\begin{equation}
\left\vert \lambda _{0},\varphi \right\rangle =\prod_{i}\left\vert \varphi
_{i}\right\rangle \otimes \left\vert \lambda _{0}\right\rangle ,  \label{GTW}
\end{equation}%
where
\begin{equation}
\left\vert \varphi _{i}\right\rangle =\left[ \cos (\frac{\pi }{4}-\frac{%
\varphi _{i}}{2}),\sin (\frac{\pi }{4}-\frac{\varphi _{i}}{2})\right] ^{T}
\label{Spin}
\end{equation}%
and
\begin{equation}
\left\vert \lambda _{0}\right\rangle =\exp \left( -\frac{\lambda _{0}^{2}}{2}%
+\lambda _{0}a^{\dag }\right) \left\vert 0\right\rangle  \label{Boson}
\end{equation}%
are the spin and boson coherent states, to describe both the
antiferromagnetic and superradiant properties.

Since the antiferromagnetic exchange interaction ($J>0$) leads to a
staggered arrangement of all spins in the $z$ direction, we should consider
two sublattices with $\varphi _{1}$ and $-\varphi _{2}$, which corresponds
to the odd and even sites of spins, respectively, in the ground-state wave
function. After a straightforward calculation, the scaled ground-state
energy
\begin{equation}
E=\left\langle \lambda _{0},\varphi _{1},\varphi _{2}\right\vert H\left\vert
\lambda _{0},\varphi _{1},\varphi _{2}\right\rangle /N  \label{SGT}
\end{equation}%
is given by
\begin{equation}
E=\omega \lambda ^{2}-J\sin \varphi _{1}\sin \varphi _{2}+\frac{\varepsilon
}{2}\left( \sin \varphi _{1}-\sin \varphi _{2}\right) +g\lambda \left( \cos
\varphi _{1}+\cos \varphi _{2}\right) ,  \label{GSEN}
\end{equation}%
where $\lambda =\lambda _{0}/\sqrt{N}$ and the parameters ($\lambda $, $%
\varphi _{1}$ and $\varphi _{2}$) are to be determined.

As shown in the Methods section, by minimizing the ground-state energy $E$
with respect to the variational parameters $(\lambda ,\varphi _{1},\varphi
_{2})$, we obtain three equilibrium equations:
\begin{equation}
\omega \lambda +g\left( \cos \alpha \cos \beta \right) =0,  \label{EQ1}
\end{equation}%
\begin{equation}
J\left( \sin 2\alpha -\sin 2\beta \right) -\varepsilon \cos (\alpha +\beta
)-2g\lambda \sin (\alpha +\beta )=0,  \label{EQ2}
\end{equation}%
\begin{equation}
J\left( \sin 2\alpha -\sin 2\beta \right) +\varepsilon \cos (\alpha -\beta
)-2g\lambda \sin (\alpha -\beta )=0,  \label{EQ3}
\end{equation}%
where
\begin{equation}
\alpha =\frac{1}{2}(\varphi _{1}+\varphi _{2})\in \left[ -\frac{\pi }{2},%
\frac{\pi }{2}\right] ,  \label{Alpha}
\end{equation}%
\begin{equation}
\beta =\frac{1}{2}(\varphi _{1}-\varphi _{2})\in \left[ -\frac{\pi }{2},%
\frac{\pi }{2}\right] .  \label{Beita}
\end{equation}%
\ These equilibrium equations, together with the stable conditions (see the
Methods section), determine the ground-state energy in Eq. (\ref{GSEN}) and
the order parameters, such as the mean-photon number $\left\langle a^{\dag
}a\right\rangle $, the magnetization $\left\langle S_{z}\right\rangle $ and
the staggered magnetization $\left\langle M_{s}\right\rangle $, which are
given respectively by
\begin{equation}
\left\{
\begin{array}{l}
\frac{\left\langle a^{\dag }a\right\rangle }{N}=\frac{g^{2}}{\omega ^{2}}%
\cos ^{2}\alpha \cos ^{2}\beta \\
\frac{\left\langle S_{z}\right\rangle }{N}=\frac{1}{N}\left\langle
\sum_{i}\sigma _{z}^{i}\right\rangle =\cos \alpha \sin \beta \\
\left\langle M_{s}\right\rangle =\frac{1}{N}\left\vert \sum_{i}\left(
-1\right) ^{i}\left\langle \sigma _{i}^{z}\right\rangle \right\vert
=\left\vert \sin \alpha \cos \beta \right\vert%
\end{array}%
\right. .  \label{OPS}
\end{equation}%
The introduction of the order parameter, the staggered magnetization $%
\left\langle M_{s}\right\rangle $, is to conveniently discuss the
antiferromagnetic properties of the Hamiltonian (\ref{FNH}). After the
ground-state energy, and especially, the order parameters, are obtained,
several rich phase diagrams can be obtained.

We first address two known limits. The first is the case when $J=0$, in
which the Hamiltonian (\ref{FNH}) reduces to the standard Dicke model \cite%
{RHD}
\begin{equation}
H_{\text{D}}=\varepsilon \sum_{i}\sigma _{z}^{i}+\frac{g}{\sqrt{N}}%
\sum_{i}\sigma _{x}^{i}\left( a+a^{\dag }\right) +\omega a^{\dag }a.
\label{DH}
\end{equation}%
By means of the equilibrium equations (\ref{EQ1})-(\ref{EQ3}) and the stable
conditions, we find
\begin{equation}
\alpha =0,\beta =-\frac{\pi }{2}  \label{R1}
\end{equation}%
for $g<g_{c}$ and
\begin{equation}
\alpha =0,\beta =-\arcsin \left[ \frac{\varepsilon }{2(J+g^{2}/\omega )}%
\right]  \label{R2}
\end{equation}%
for $g>g_{c}$, where $g_{c}=\sqrt{\varepsilon \omega /2}$, i.e.,
\begin{equation}
\frac{\left\langle a^{\dag }a\right\rangle }{N}=0,\frac{\left\langle
S_{z}\right\rangle }{N}=-1  \label{R3}
\end{equation}%
for $g<g_{c}$ and
\begin{equation}
\frac{\left\langle a^{\dag }a\right\rangle }{N}=\frac{g^{2}}{\omega ^{2}}-%
\frac{\varepsilon ^{2}}{4g^{2}},\frac{\left\langle S_{z}\right\rangle }{N}=-%
\frac{\varepsilon \omega }{2g^{2}}  \label{R4}
\end{equation}%
for $g>g_{c}$. This means that a second-order quantum phase transition from
the normal phase ($g<g_{c}$) to the superradiant phase ($g>g_{c}$) occurs
\cite{CE03,CG08}, as shown in Fig. 2(a). Moreover, the Dicke model has $U(1)$%
\ symmetry in the normal phase. Whereas, in the superradiant phase the
system acquires macroscopic collective excitations governed mainly by the
collective spin-photon interaction term $g\sum_{i}\sigma _{x}^{i}\left(
a+a^{\dag }\right) $, and thus it has $Z_{2}^{z}\otimes Z_{2}$\ symmetry,
where $Z_{2}^{z}$\ is the global rotation of $\pi $ around the $z$\ axis
\cite{TK92} and $Z_{2}$ is the change of sign of the boson coherent state ($%
\left\vert \lambda _{0}\right\rangle \rightarrow -\left\vert \lambda
_{0}\right\rangle $). In experiments, this quantum phase transition has been
observed \cite{KB10,KB11,RH13} in an optical cavity with a Bose-Einstein
condensate by measuring the mean-photon number $\left\langle a^{\dag
}a\right\rangle $ and the magnetization $\left\langle S_{z}\right\rangle /N$%
. Recently, it has been well investigated in many-body circuit QED \cite%
{GC07,NL09,PN10,HI12,PN11,OV11} and spin-orbit-driven Bose-Einstein
condensate \cite{YP13}.

For $g=0$, the Hamiltonian (\ref{FNH}) turns into the Ising model \cite%
{AAO03}
\begin{equation}
H_{\text{I}}=J\sum_{i}\sigma _{z}^{i}\sigma _{z}^{i+1}+\varepsilon
\sum_{i}\sigma _{z}^{i},  \label{Ising}
\end{equation}%
in which a first-order phase transition from the paramagnetic phase to the
antiferromagnetic phase at the critical point $J_{c}=\varepsilon /2$ can be
recovered. In the paramagnetic phase ($J<J_{c}$), the ground-state wave
function is $\left\vert \ldots \downarrow \downarrow \downarrow \downarrow
\downarrow \downarrow \downarrow \downarrow \ldots \right\rangle $, which
implies that the system has translation symmetry and $\left\langle
S_{z}\right\rangle /N=-1$ and $\left\langle M_{s}\right\rangle =0$. In the
antiferromagnetic phase ($J>J_{c}$), the ground-state wave function becomes $%
\left\vert \ldots \uparrow \downarrow \uparrow \downarrow \uparrow
\downarrow \uparrow \downarrow \ldots \right\rangle $, in which translation
symmetry is broken and $\left\langle S_{z}\right\rangle /N=0$ and $%
\left\langle M_{s}\right\rangle =1$.

If both $g$ and $J$ are non-zeros, we find four different regions: (i) $%
\alpha =0$, $\beta =-\pi /2$, (ii) $\alpha =+\pi /2$, $\beta =0$, (iii) $%
\alpha =0$, $\beta \neq 0$, and (iv) $\alpha \neq 0$, $\beta \neq 0$.
Specially, the order parameters in these four regions are given respectively
by
\begin{equation}
\text{(i) }\frac{\left\langle a^{\dag }a\right\rangle }{N}=0,\frac{%
\left\langle S_{z}\right\rangle }{N}=-1,\left\langle M_{s}\right\rangle =0,
\label{I}
\end{equation}%
\begin{equation}
\text{(ii) }\frac{\left\langle a^{\dag }a\right\rangle }{N}=0,\frac{%
\left\langle S_{z}\right\rangle }{N}=0,\left\langle M_{s}\right\rangle =1,
\label{II}
\end{equation}%
\begin{equation}
\text{(iii) }\frac{\left\langle a^{\dag }a\right\rangle }{N}=\frac{g^{2}}{%
\omega ^{2}}\left[ 1-\frac{\varepsilon ^{2}}{4(J+\frac{g^{2}}{\omega })^{2}}%
\right] ,\frac{\left\langle S_{z}\right\rangle }{N}=-\frac{\varepsilon }{2(J+%
\frac{g^{2}}{\omega })},\left\langle M_{s}\right\rangle =0,  \label{III}
\end{equation}%
\begin{equation}
\text{(iv) }\frac{\left\langle a^{\dag }a\right\rangle }{N}=\frac{g^{2}}{%
\omega ^{2}}\cos ^{2}\alpha \cos ^{2}\beta ,\frac{\left\langle
S_{z}\right\rangle }{N}=\cos \alpha \sin \beta ,\left\langle
M_{s}\right\rangle =\left\vert \sin \alpha \cos \beta \right\vert ,
\label{IV}
\end{equation}%
where
\begin{equation}
\cos \alpha =\frac{-\varepsilon \sin \beta }{2\left( J-g^{2}\cos ^{2}\beta
/\omega \right) },  \label{a}
\end{equation}%
\begin{equation}
\cos ^{2}\beta =\frac{\omega }{g^{2}}\left[ J-\frac{\varepsilon }{2}\sqrt{1-%
\frac{g^{2}}{\omega J}}\right] .  \label{b}
\end{equation}

In terms of the different properties of the order parameters, the cases
(i)-(iv) are denoted by PNP, ANP, PSP, and ASP, respectively. The ANP ($%
\left\langle a^{\dag }a\right\rangle /N=0$, $\left\langle S_{z}\right\rangle
/N=0$ and $\left\langle M_{s}\right\rangle =1$) and the ASP ($\left\langle
a^{\dag }a\right\rangle /N\neq 0$, $\left\langle S_{z}\right\rangle /N\neq
-1 $ and $\left\langle M_{s}\right\rangle \neq 1$) are new phases in
many-body quantum optics. In Fig. 2, we plot phase diagrams for different
antiferromagnetic spin-spin nearest-neighbor interactions. This figure shows
clearly that these predicted quantum phases can be driven by the collective
spin-photon coupling strength $g$, the antiferromagnetic nearest-neighbor
spin-spin interaction $J$, and the effective magnetic field $\varepsilon $.
Especially, the region of the ASP becomes larger when increasing $J$.

In Fig. 3, we plot phase diagrams as functions of the antiferromagnetic
nearest-neighbor spin-spin interaction $J$ and the collective spin-photon
coupling strength $g$ for different effective magnetic fields (a) $%
\varepsilon =0$ and (b) $\varepsilon =\omega /4$. In the absence of $%
\varepsilon $, Eq. (\ref{FNH}) reduces to the form
\begin{equation}
\mathbf{\ }H=J\sum_{i}\sigma _{z}^{i}\sigma _{z}^{i+1}+\frac{g}{\sqrt{N}}%
\sum_{i}\sigma _{x}^{i}\left( a+a^{\dag }\right) +\omega a^{\dag }a.
\label{NH}
\end{equation}%
In such a case, only the ANP and the PSP can be found, as shown in Fig.
3(a). When increasing $\varepsilon $, four quantum phases are predicted
again, as shown in Fig. 3(b). In addition, by means of the ground-state
energy, we find that all transitions between these different quantum phases
in Figs. 2 and 3 are of second order.

\textbf{Symmetry.} In order to better understand these predicted quantum
phases, it is necessary to discuss the corresponding symmetries. For the PNP
and the PSP, the system properties are similar to those of the normal phase
and the superradiant phase in the standard Dicke model, i.e., the system
displays both $U(1)$ and translation symmetries in the PNP, and becomes $%
Z_{2}^{z}\otimes Z_{2}$ and translation symmetries in the PSP. However, in
the ANP, though no photon is excited, the antiferromagnetic order emerges.
This implies that in such a case only $U(1)$ symmetry can be found.
Interestingly, in the ASP the Hamiltonian (\ref{FNH}) is governed mainly by
the term
\begin{equation}
H_{gJ}=J\sum_{i}\sigma _{z}^{i}\sigma _{z}^{i+1}+\ g\sum_{i}\sigma
_{x}^{i}\left( a+a^{\dag }\right) /\sqrt{N},  \label{GJ}
\end{equation}%
in which there exists a competition between the antiferromagnetic
nearest-neighbor spin-spin interaction and the collective spin-photon
interaction. As a result, both the antiferromagnetic and superradiant orders
coexist and the system possesses $Z_{2}^{z}\otimes Z_{2}$\ symmetry, i.e.,
both $U(1)$ and translation symmetries are broken simultaneously.

\textbf{Possible experimental observation.} We first estimate the parameters
for experiments. When we choose $C_{\Sigma }\sim 600$\ aF, $C\sim 20$\ aF, $%
E_{J}^{0}\sim 2\pi \times 3.5$\ GHz,\ $S_{0}\sim 1$\ $\mu $m$^{2}$, $d\sim
10 $\ $\mu $m, $L_{0}\sim 19$\ mm, $\omega \sim 2\pi \times 6.729$\ GHz, and
$N=100$ \cite{JMF09}, the antiferromagnetic nearest-neighbor spin-spin
interaction parameter and the collective spin-photon coupling strength are
given respectively by $J\sim 2\pi \times 2.2$\ GHz and $g\sim 2\pi \times
1.5 $\ GHz ($g_{0}=0.01$ is responsible for the Lamb-Dicke approximation).
In addition, the effective magnetic field $\varepsilon $ can range from $0$
to $2\pi \times 6.8$ GHz by controlling the gate voltage $V_{g}$. These
parameters ensure that the system should probe the predicted phase
transitions.\ To observe these phase transitions, the relaxation time $T_{1}$%
\ and the coherence time $T_{2}$\ should be much smaller than the lifetime $%
1/\kappa $ of the photon, i.e., $T_{1}$\ $>T_{2}>1/\kappa =23.4$\ ns, where $%
\kappa $ is the decay rate of the photon. This restriction can be easy to
satisfy in current experimental setups (for example, $T_{1}=7.3$\ $\mu $s
and $T_{2}$\ $=500$\ ns in Ref. \cite{WA2005}).

We now illustrate how to identify these different quantum phases. Here we
propose to detect four phases by measuring both the mean-photon number $%
\left\langle a^{\dag }a\right\rangle $ and the magnetization $\left\langle
S_{z}\right\rangle $. For the PNP and the ANP, we can separate these by
directly observing the magnetization because $\left\langle
S_{z}\right\rangle /N=-1$ in the PNP and $\left\langle S_{z}\right\rangle
/N=0$ in the ANP, as shown in Fig. 4(a). For the PSP and the ASP, both the
photon and the spin are collectively excited. Moreover, when increasing the
effective magnetic field $\varepsilon $, $\left\langle S_{z}\right\rangle /N$
always decreases. This means that it is difficult to distinguish the PSP and
the ASP by measuring $\left\langle S_{z}\right\rangle /N$. Fortunately, we
find that in the ASP the mean-photon number has an unconventional behavior
that could increase it from zero to a finite value and then decrease, as
shown in Fig. 4(b). The relevant physics can be understood as follows. When
the effective magnetic field $\varepsilon $\ is applied, it can initially
promote the arrangement of all spins from the antiferromagnetic to the
paramagnetic terms \cite{JS11}. For example, in the case of a weak $%
\varepsilon $, the spin arrangement becomes $\left\vert \ldots \uparrow
\downarrow \uparrow \{\downarrow \downarrow \downarrow \}\uparrow \downarrow
\uparrow \ldots \right\rangle $ from the antiferromagnetic case $\left\vert
\ldots \uparrow \downarrow \uparrow \downarrow \uparrow \downarrow \uparrow
\downarrow \ldots \right\rangle $. This process is helpful for achieving
photon-induced collective excitations. Thus, the mean-photon number can
increase. However, the rearrangement of spins gives rise to an opposite
result of the magnetization, i.e., it decreases when increasing $\varepsilon
$. For strong $\varepsilon $, this effective magnetic field in the $z$ axis
leads to a large spin imbalance $\left\vert \ldots \uparrow \uparrow
\downarrow \downarrow \downarrow \downarrow \downarrow \downarrow \ldots
\right\rangle $ and thus suppresses the spin-photon collective excitations,
i.e., both $\left\langle a^{\dag }a\right\rangle /N$ and $\left\langle
S_{z}\right\rangle /N$ decrease when increasing $\varepsilon $. In terms of
the different behaviors of both $\left\langle a^{\dag }a\right\rangle /N$
and $\left\langle S_{z}\right\rangle /N$, we argue that our predicted
quantum phases can be identified.

It should be pointed out that the microwave photon in superconducting
circuits is not easy to measure directly \cite{Nation} via photon-number
detectors, because its energy ($\hbar \omega $) is very small. However, in
the dispersive region $\Delta \gg g$, where $\Delta =\varepsilon -\omega $,
the photon number can be detected by the photon-number-dependent light shift
(the Stark plus Lamb shifts) of the atom transition frequency \cite{Schuster}%
. Unfortunately, to achieve the predicted phase diagrams, the system should
work at the quasi-dispersive-strong region $\Delta /g<4$. In such a region,
the above approach to detect photons does not work. Recently, it has been
proposed \cite{Rom,Chen} to detect the photon by irreversible absorption of
photons. In these proposals \cite{Rom,Chen}, the absorbers along the
waveguide are built with bistable quantum circuits, and can produce a large
voltage pulse when the photon decays into a stable state. This suggests that
in future experiments the mean-photon number could be detected, and then our
predicted phase diagrams could also be observed. \newline

{\LARGE \textbf{Discussion}}

Let us here address the no-go theorem in quantum optics\textbf{.} This no-go
theorem, demonstrated \cite{RK75} in 1975, shows that in a typical optical
cavity with an ensemble of natural two-level atoms, the phase transition
from the normal phase to the superradiant phase is forbidden by the $A^{2}$
term, where $A$ is the vector potential. Recently, the no-go theorem has
been addressed \cite{PN10,OV11} in circuit QED, with many superconducting
qubits interacting with a quantized voltage (microwave photon). However, in
this report, the required microwave photon is generated from the
quantization of the magnetic flux. In such a case, no $A^{2}$ term can
occur, i.e., the no-go theorem is not valid.

We now consider how the decay of the photon and the disorder in fabrication
affect the predicted phase diagrams. When considering the decay of the
photon, the stationary mean-photon number, which can be derived from
\begin{equation}
i\frac{\partial a}{\partial t}=\omega a+\frac{g}{\sqrt{N}}S_{x}-i\kappa a=0,
\label{HSP}
\end{equation}%
becomes
\begin{equation}
\left\langle a^{\dag }a\right\rangle =\frac{g^{2}}{N(\omega ^{2}+\kappa ^{2})%
}\left\langle S_{x}\right\rangle ^{2},  \label{MPA}
\end{equation}%
whereas the stationary value of $\left\langle S_{x}\right\rangle $ remains
unchanged (i.e., it is identical to the case without photon decay). This
means that we can use an effective Hamiltonian
\begin{equation}
H_{\text{eff}}=J\sum_{i}\sigma _{z}^{i}\sigma _{z}^{i+1}+\frac{g}{\sqrt{N}}%
\sum_{i}\sigma _{x}^{i}\left( a+a^{\dag }\right) +\omega _{\text{eff}%
}a^{\dag }a,  \label{HEFF}
\end{equation}%
with
\begin{equation}
\omega _{\text{eff}}=\sqrt{\omega ^{2}+\kappa ^{2}},  \label{Omee}
\end{equation}%
to discuss the phase diagrams \cite{RH13} induced by the decay of the
photon. Since the decay rate $\kappa $ of the photon ($\sim $ MHz) is far
smaller than the other parameters ($\sim $ GHz), the decay of the photon has
almost no effect on the predicted phase diagrams.

In addition, the imperfections in fabrication result in a weak randomness in
the antiferromagnetic nearest-neighbor spin-spin interaction \cite{AAH12}.
Moreover, the disorder antiferromagnetic interaction generates a disordered
phase, such as a random singlet phase \cite{Fisher}. In this phase, most
spins form a singlet pair with nearby spins, and the residual induce weak
long-distance pairs. Unfortunately, the disordered phase is only a local
correlation, and is thus unstable in the presence of a strong external
magnetic field (i.e., for a weak magnetic field, the disordered phase can
occur). The phase diagrams predicted here should be observable under a
strong magnetic field (See Figs. 2(c) and 2(d)). This means that these
predicted quantum phases will not be qualitatively affected by weak disorder.

Mean-field predictions become more accurate for larger number of spins.
However, mean-field is often a good starting point, and provides some basic
insight in the system. Moreover, for current experimental techniques, the
spin number is not sufficiently large, but this should change in the future.
For smaller number of spins, we can perform direct numerical diagonalization
to discuss the ground-state properties. In Fig. 5, we plot the order
parameters $\left\langle a^{\dag }a\right\rangle /N$, $\left\langle
S_{z}\right\rangle /N$, and $\left\langle M_{s}\right\rangle $\textbf{\ }as
functions of the collective spin-photon coupling strength $g$ and the
effective magnetic field $\varepsilon $. This result shows clearly that for
a small number of spins, the predicted quantum phases still exist, but the
phase boundaries are affected significantly.

In summary, we have investigated the Dicke-Ising model with an
antiferromagnetic nearest-neighbor spin-spin interaction in circuit QED for
a superconducting qubit array and predicted four quantum phases, including
the PNP, the ANP, the PSP and the ASP, with different symmetries. Moreover,
all transitions between these different quantum phases are of second order.
We have also found an unconventional photon signature in the ASP, where both
the antiferromagnetic and superradiant orders coexist. We believe that this
system allows to explore exotic many-body physics in quantum optics and
condensed-matter physics because it has an interesting competition between
the collective spin-photon interaction and the nearest-neighbor spin-spin
interaction. \newline

{\LARGE \textbf{Methods}}

\textbf{Three equilibrium equations and the corresponding stable conditions.}
Here we present detailed calculations on how to derive the three equilibrium
equations (\ref{EQ1})-(\ref{EQ3}) and the corresponding stable conditions.
After minimizing the ground-state energy $E$ in Eq. (\ref{GSEN}) with
respect to the variational parameters ($\lambda $, $\varphi _{1}$ and $%
\varphi _{2}$), we obtain three equations:
\begin{equation}
\frac{\partial E}{\partial \lambda }=2\omega \lambda +g\left( \cos \varphi
_{1}+\cos \varphi _{2}\right) =0,  \label{EL}
\end{equation}%
\begin{equation}
\frac{\partial E}{\partial \varphi _{1}}=-J\cos \varphi _{1}\sin \varphi
_{2}+\frac{\varepsilon }{2}\cos \varphi _{1}-g\lambda \sin \varphi _{1}=0,
\label{EP1}
\end{equation}%
\begin{equation}
\frac{\partial E}{\partial \varphi _{2}}=-J\sin \varphi _{1}\cos \varphi
_{2}+\frac{\varepsilon }{2}\cos \varphi _{2}-g\lambda \sin \varphi _{2}=0.
\label{EP2}
\end{equation}%
After defining $\alpha =(\varphi _{1}+\varphi _{2})/2\in \left[ -\pi /2,\pi
/2\right] $ and $\beta =(\varphi _{1}-\varphi _{2})/2\in \left[ -\pi /2,\pi
/2\right] $, the three equilibrium equations (\ref{EQ1})-(\ref{EQ3}) are
easily derived.

In addition, by means of three parameters ($\lambda $, $\alpha $ and $\beta $%
), the ground-state stability is determined by the $3\times 3$ Hessian matrix%
\begin{equation}
\mathcal{M}\!\!\!=\left[
\begin{array}{lll}
\frac{\partial ^{2}E}{\partial \lambda ^{2}} & \frac{\partial ^{2}E}{%
\partial \lambda \partial \alpha } & \frac{\partial ^{2}E}{\partial \lambda
\partial \beta } \\
\frac{\partial ^{2}E}{\partial \alpha \partial \lambda } & \frac{\partial
^{2}E}{\partial \alpha ^{2}} & \frac{\partial ^{2}E}{\partial \alpha
\partial \beta } \\
\frac{\partial ^{2}E}{\partial \beta \partial \lambda } & \frac{\partial
^{2}E}{\partial \beta \partial \alpha } & \frac{\partial ^{2}E}{\partial
\beta ^{2}}%
\end{array}%
\right] .  \label{HESS}
\end{equation}%
If $\mathcal{M}$ is positive definite (i.e., all eigenvalues $h_{i}$ of $%
\mathcal{M}$ are positive), the system is located at the stable phases. If $%
\mathcal{M}$ is negative definite (i.e., all eigenvalues $h_{i}$ of $%
\mathcal{M}$ are negative), the system is unstable.


\begin{addendum}

 \item  TJ is supported by the 973 Program under Grant No. 2012CB921603, the NNSFC under Grant No. 61378015, the International Science and Technology Cooperation Program of China under
Grant No. 2011DFA12490, and the PCSIRT under Grant No. IRT13076. JL is supported by the NNSFC under grant No. 11275118. GC is supported by the NNSFC under Grant No. 61275211, the NCET under Grant No. 13-0882, the FANEDD under Grant No. 201316, and the OIT under Grant No. 2013804. LY is supported by the ZJNSF under Grant No. LY13A040001 and the ZJSRFED under Grant No. Y201122352. FN is partially supported by the RIKEN iTHES Project, MURI Center for Dynamic Magneto-Optics, JSPS-RFBR contract No. 12-02-92100, Grant-in-Aid for Scientific Research (S), MEXT Kakenhi on Quantum Cybernetics, and the JSPS via its FIRST program.

 \item[Author Contributions] G.C., S.J. and F.N. conceived the idea. Y.Z, L.Y., and J.-Q.L. performed the calculation. G.C., S.J. and F.N. wrote the manuscript.

 \item[Competing Interests] The authors declare that they have no
competing financial interests.

\end{addendum}
\ifthenelse{\boolean{SubmittedVersion}}{\processdelayedfloats}{%
\cleardoublepage}


\textbf{Figure 1: Proposed circuit QED with a superconducting qubit array.}

\textbf{Figure 2: Phase diagrams for different antiferromagnetic spin-spin
nearest-neighbor interactions.} The plotted parameters are chosen as (a) $%
J/\omega =0$, (b) $J/\omega =0.05$, (c) $J/\omega =0.1$, and (d) $J/\omega
=0.2$, respectively. In (b), (c) and (d), the phase boundaries (from top to
bottom) are determined respectively by $\varepsilon _{c}=2(J+g^{2}/\omega )$%
, $\varepsilon _{c}=2(J+g^{2}/\omega )\sqrt{1-g^{2}/(\omega J)}$, and\textbf{%
\ }$\varepsilon _{c}=2J\sqrt{1-g^{2}/(\omega J)}$.

\textbf{Figure 3: Phase diagrams for different effective magnetic fields.}
The plotted parameters are chosen as (a) $\varepsilon /\omega =0$ and (b) $%
\varepsilon /\omega =0.25$, respectively. In (a), the phase boundary is
determined by $J_{c}=g^{2}/\omega $. In (b), the phase boundaries (from top
to bottom) are given respectively by $J_{c}=(g^{2}/\omega +\sqrt{\varepsilon
^{2}+g^{4}/\omega ^{2}})/2$,\textbf{\ }$(J_{c}+g^{2}/\omega )\sqrt{%
1-g^{2}/(\omega J_{c})}=\varepsilon /2$, and\textbf{\ } $J_{c}=\varepsilon
/2-g^{2}/\omega $.

\textbf{Figure 4: Order parameters versus the effective magnetic field.} In
(a) and (b), the magnetization $\left\langle S_{z}\right\rangle /N$ and the
mean-photon number $\left\langle a^{\dag }a\right\rangle /N$ as functions of
the effective magnetic field $\varepsilon $ are plotted, respectively, when $%
J/\omega =0.30$ and $g/\omega =0.25$.

\textbf{Figure 5: Order parameters via the direct numerical diagonalization
method.} In (a)-(c), the mean-photon number $\left\langle a^{\dag
}a\right\rangle /N$, the magnetization $\left\langle S_{z}\right\rangle /N$,
and the the staggered magnetization $\left\langle M_{s}\right\rangle $ as
functions of the collective spin-photon coupling strength $g$ and the
effective magnetic field $\varepsilon $ are plotted, respectively, when $%
J/\omega =0.1$ and $N=7$.


\clearpage

\begin{figure}[t]
\centering\includegraphics[width = 0.7\linewidth]{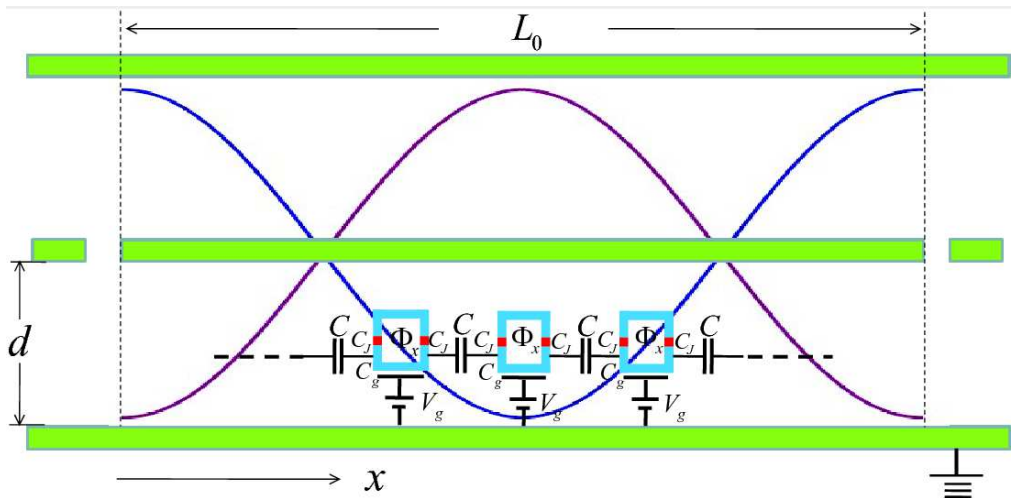}
\caption{\textbf{Proposed circuit QED with a superconducting qubit array.}}
\label{fig1}
\end{figure}

\begin{figure}[t]
\centering\includegraphics[width = 0.8\linewidth]{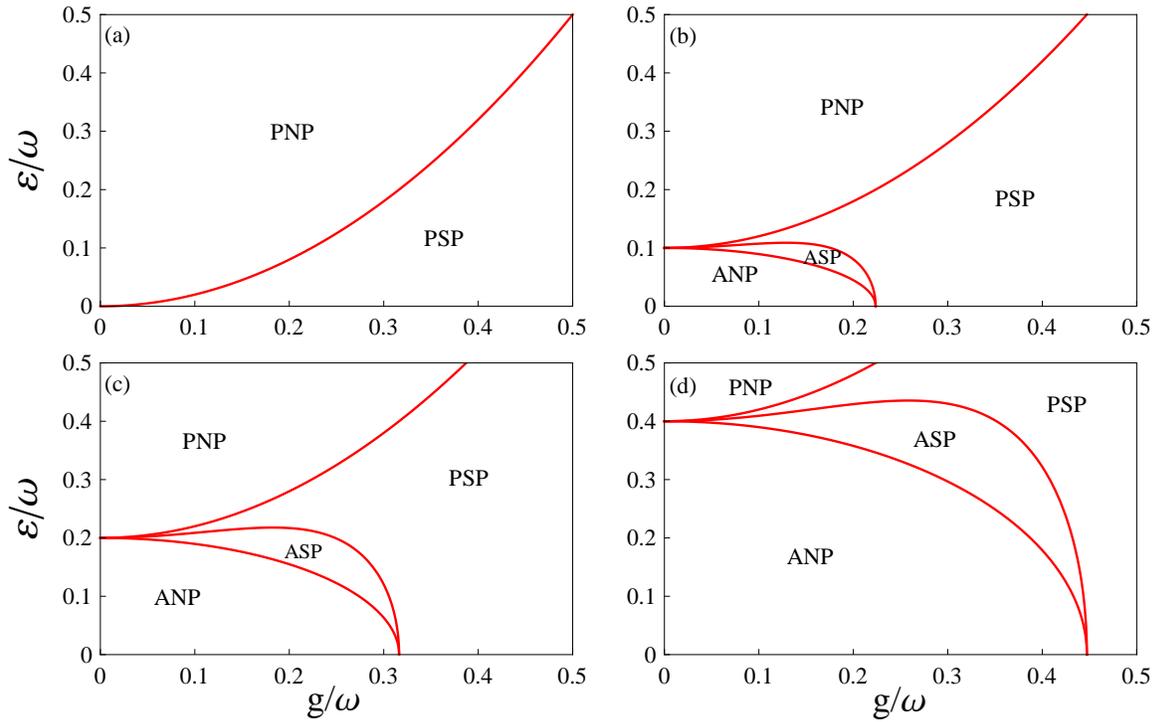}
\caption{\textbf{Phase diagrams for different antiferromagnetic spin-spin
nearest-neighbor interactions.}}
\label{fig2}
\end{figure}

\begin{figure}[t]
\centering\includegraphics[width = 0.8\linewidth]{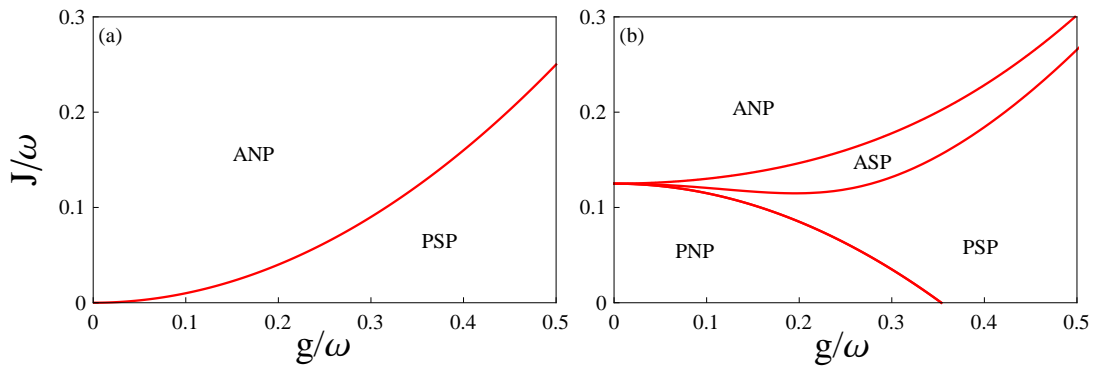}
\caption{\textbf{Phase diagrams for different effective magnetic fields.}}
\label{fig3}
\end{figure}

\begin{figure}[t]
\centering\includegraphics[width = 0.7\linewidth]{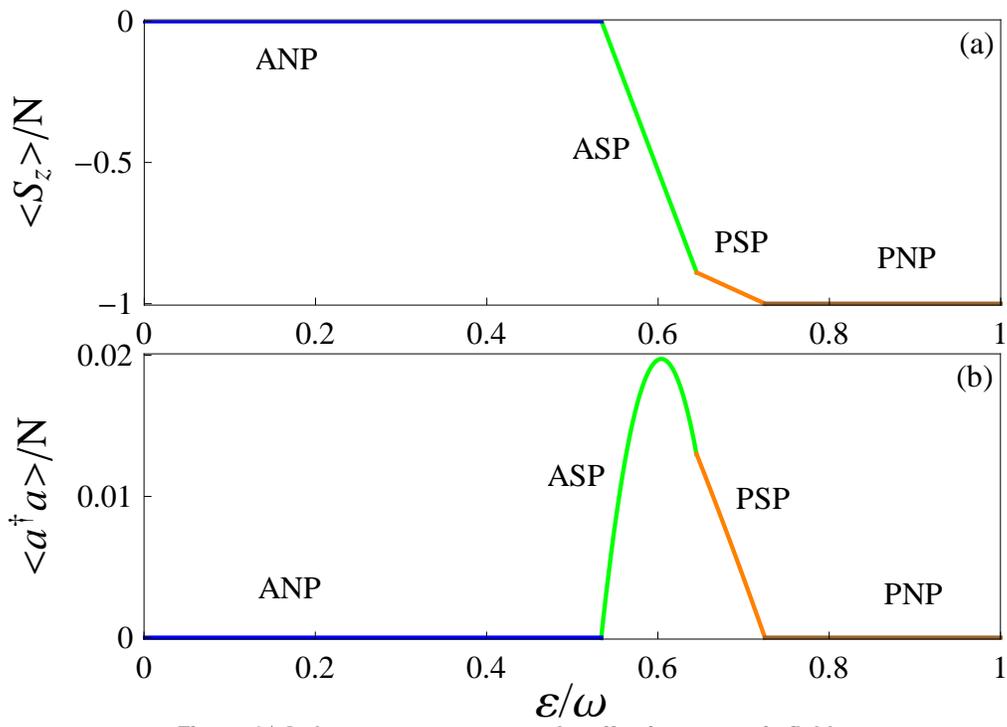}
\caption{\textbf{Order parameters versus the effective magnetic field.}}
\label{fig4}
\end{figure}

\begin{figure}[t]
\centering\includegraphics[width = 0.7\linewidth]{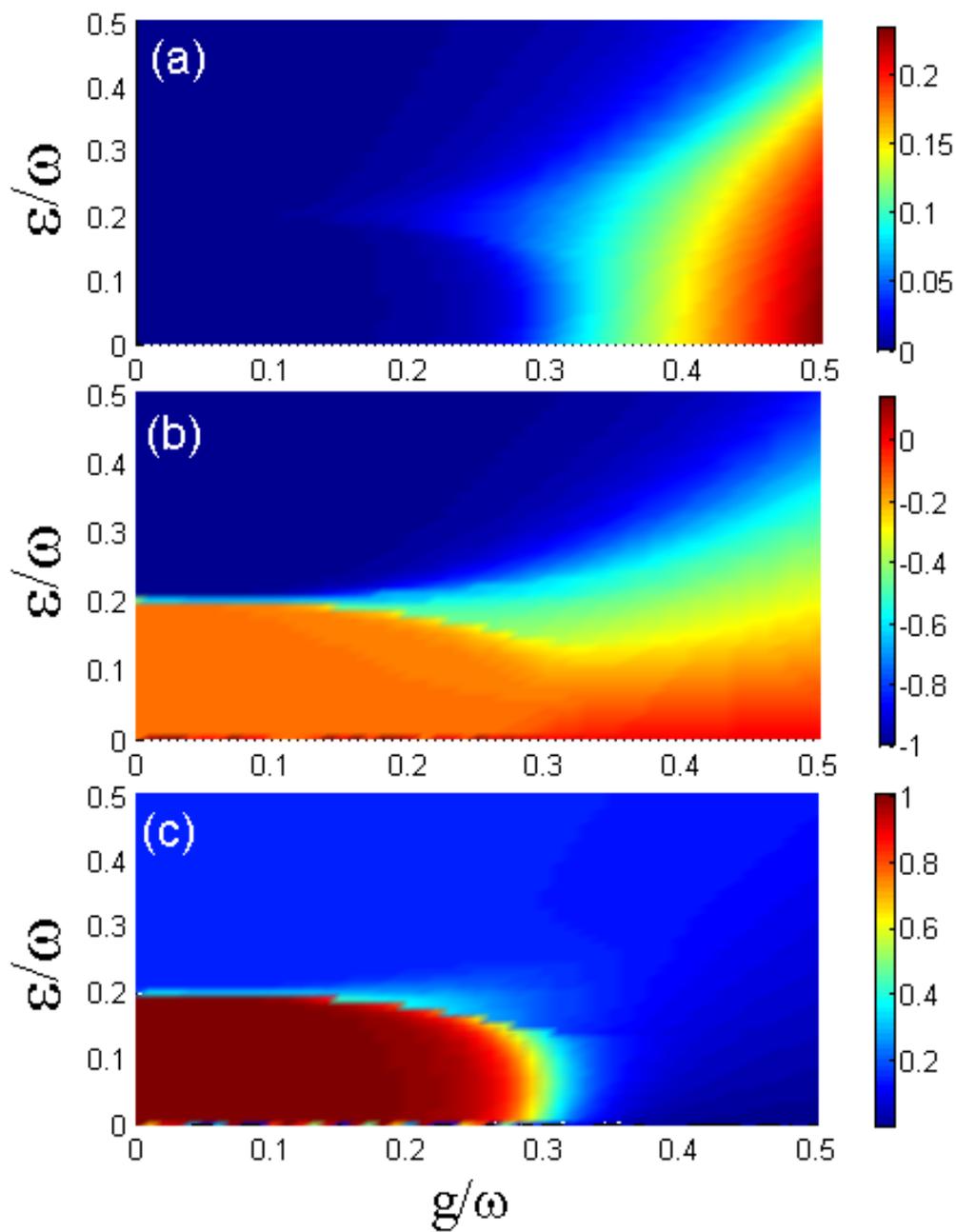}
\caption{\textbf{Order parameters via the direct numerical diagonalization
method.}}
\label{fig5}
\end{figure}

\end{document}